\documentstyle[11pt]{article}
\begin{document}


\def\e{{\hat e}}
\def\ie{{$i.e.$}}
\def\ra{{\rightarrow}}
\def\a{{\alpha}}
\def\b{{\beta}}
\def\eps{{\epsilon}}
\def\n{{\eta}}
\def\g{\gamma}
\def\s{{\sigma}}
\def\r{{\rho}}
\def\z{{\zeta}}
\def\x{{\xi}}
\def\d{{\delta}}
\def\t{{\theta}}
\def\T{{\Theta_A}}
\def\l{{\lambda}}
\def\L{{\Lambda}}
\def\ca{{\cal A}}
\def\cc{{\cal C}}
\def\cd{{\cal D}}
\def\ce{{\cal E}}
\def\cg{{\cal G}}
\def\ci{{\cal I}}
\def\co{{\cal O}}
\def\cn{{\cal N}}
\def\car{{\cal R}}
\def\cs{{\Theta_S}}
\def\cu{{\cal U}}
\def\cv{{\cal V}}
\def\cz{{\cal Z}}
\def\ck{{\cal K}}
\def\pr{{\partial}}
\def\prt{{\pr}_{\t}}
\def\tri{{\triangle}}
\def\na{{\nabla }}
\def\S{{\Sigma}}
\def\G{{\Gamma }}
\def\sp{\vspace{.1in}}
\def\hs{\hspace{.25in}}

\newcommand{\be}{\begin{equation}} \newcommand{\ee}{\end{equation}}
\newcommand{\bea}{\begin{eqnarray}}\newcommand{\eea}
{\end{eqnarray}}


\begin{titlepage}

\begin{flushright}
{{G\"oteborg preprint}\\{hep-th/0006073}\\
{June, 2000}}
\end{flushright}

\begin{center}
\baselineskip= 24 truept
\vspace{.5in}

{\Large \bf D-branes, Cyclic Symmetry and Noncommutative Geometry}

\vspace{.4in}
{\large{\sc Supriya Kar}\footnote{supriya@fy.chalmers.se}}

\end{center}
\begin{center}
\baselineskip= 18 truept

{\large \it Institute for Theoretical Physics \\
G\"oteborg University and Chalmers University of Technology \\
S-412 96 G\"oteborg, Sweden}

\vspace{.2in}

({\today})

\end{center}

\vspace{.4in}

\baselineskip= 15 truept

\begin{center}
{\bf Abstract}
\end{center}
\vspace{.25in}

We investigate the open string modes, describing the world-volume of a 
D p-brane, for its cyclic symmetry in presence of a magnetic field. 
It is argued that the constant coordinate modes receive non-perturbative
correction. We show that they introduce the notion of
noncommutativity on the D p-brane world-volume and make it UV-renormalizable.
An analogy between cyclic symmetry ($\a'$-corrections) and the 
noncommutative geometry ($\Theta_A$-corrections) is presented to explain
some of the unusual IR phenomena often noticed in a noncommutative theory.

\vspace{.25in}

\thispagestyle{empty}
\end{titlepage}

\baselineskip= 18 truept

\section{Introduction}

Noncommutative geometry has renewed its interest in a
non-perturbative formulation \cite{polchinski}, 
stimulated with the work \cite{connes}
by Connes, Douglas and Schwarz. It is shown that
M-theory in a constant three-form background is equivalent to a
super Yang-Mills theory on a noncommutative torus. The interest has
further gained momentum with the work \cite{seiberg} by Seiberg and Witten 
where an equivalence between Yang-Mills symmetry and
a noncommutative gauge symmetry has been established. Since the
theory includes gravity, one of the important problem is the consistency of a 
noncommutative theory or its renormalizability. Interestingly, a great deal
of work describing loop-renormalizations in a noncommutative theory has 
been discussed in the current literatures \cite{martin}-\cite{armoni}. It
is known that a noncommutative description shares some common features with
that of an open string theory \cite{witten,gross}.

\sp

In this context, we shall present a comparative study between the stringy 
($\a'$) and a noncommutative ($\Theta_A$) corrections to
an ordinary quantum field theory (QFT). Under either correction,
the space is described by quantum operators though they
possess very different geometries. The $\a'$ correction
are perturbative and introduces the notion of a string of finite length. On
the other hand, we shall note that the $\Theta_A$ gives rise to
the geometric corrections and also can be seen to introduce a minimum
length scale in the theory. In fact, both the symmetries are
intriguing to understand the ultra-violet (UV) and infra-red (IR) 
renormalizations in the respective theories.

\sp

Since a Dirichlet (D) p-brane \cite{polchinski}
is defined at the boundary of an open
string, it retains the cyclic symmetry on its world-volume. 
In presence of a magnetic field ($i.e.$ an antisymmetric two form B-field),
the D p-brane world-volume possesses a noncommutative symmetry.
Then, the world-volume turns out to be an ordered space \cite{seiberg}.
Thus, both the cyclic and noncommutative symmetry, independently, lead
to an ordered space description on the D p-brane world-volume. 
In fact, it is the conjugate momentum on the world-volume of a D p-brane 
which receives correction due to B-field and introduces the notion of
noncommutativity. As a result, the string coupling $g_s$ can be seen to
be associated with the noncommutative parameter. With a simple analysis,
we note that the geometric correction to the world-volume coordinates
is non-perturbative nature. Interestingly,  both $\a'$ and $\Theta_A$ 
corrections can be seen to contribute identical phase factors{\footnote{The
phase can be seen to be symmetric with the cyclic ordering and is
antisymmetric in case of the noncommutative symmetry.
It is the antisymmetric phase, which
turns out to be non-trivial where as the symmetric one does not contribute
substantially with the on-shell condition. 
In fact, the latter is a requirement for the cyclic symmetry.}
to the interacting vertices (or amplitude). 
In addition, with $\a'$-correction, a
T-duality symmetry is known to relate $s$ and $t$-channel amplitudes though
very different phenomena known to occur in either channels 
\cite{gsw}. Similarly,
some of the divergences in an ordinary theory can be seen to be regularized
with the geometric correction. Thus, it might be possible to explain
some of the important phenomena in a noncommutative theory with the 
tools from the stringy corrections.

\sp

In this paper, we investigate the cyclic and noncommutative symmetries
on the world-volume of a D p-brane in an open bosonic string theory. 
We consider a consistent Fourier mode expansion for the world-volume 
coordinates in presence of a magnetic field and obtain the equal-time
commutation relation for the zero modes as well as for the non-zero modes. 
Then, the constant coordinate modes turn out to be noncommutative
where as the non-zero momentum modes are modified due to the magnetic field.
The latter observation is due to the fact that the world-volume metric
receives correction in a noncommutative theory. The analysis for the zero
modes suggests that the
center of mass coordinates for a D p-brane become
operators and are completely responsible for the noncommutative description.
Then, the stringy analysis for the noncommutavity on the world-volume
reduces precisely to that of a quantum field theory. We compute the phase
factor for a set of scalar fields in a momentum space of the noncommutative
quantum field theory (NQFT). We analyze the non-trivial phase 
contribution to exploit some of the divergence problems of an ordinary quantum
field theory. We perform the analysis in open-string channel as
well as its dual closed string channel and explain some of the IR
phenomena observed \cite{minwalla} in a noncommutative theory. 

\sp

We plan to present the paper as follows. In section 2, we briefly re-view 
some of the D p-brane features in an open string channel. We discuss the
noncommutativity among the zero modes in section 3.1 and the compute for 
the non-local phase factor in 3.2. The field theoretic limit
is discussed in section 3.3. The section 4 deals with the UV/IR connection,
where in 4.1 we discuss the cyclic ordering on a D p-brane world-volume 
in presence of a magnetic field. We explain some of the IR-phenomena due
to zero modes in sections 4.2 and 4.3. Finally, we conclude
with some out-look in section 5.

\section{Preliminaries}

\subsection{D p-brane world-volume and cyclic symmetry}

Let us begin with an infinite-strip world-sheet topology, 
$\ci_{[0,\pi ]}\times \car$, for an open string.
Then the space-like coordinate, $0\leq\s\leq\pi$, corresponds to the
length of the string where as $\tau$ is time-like. 
The mode expansion for the string coordinates, $X^{\mu}(\s,\tau )$, 
in term of its zero  modes (coordinate: $x^{\mu}$ and momentum
$p^{\mu}$) and non-zero modes ( momentum: $\a^{\mu}_m$) 
can be given by
\be
X^{\mu}(\s,\tau )\ 
=\ x^{\mu}\ +\ 2\a'\ p^{\mu}\ \tau \
+\ i {\sqrt{2\a'}}\
\sum_{m\neq 0}{1\over{m}}\ \a^{\mu}_m\ e^{-i m\tau}\ \cos m\s \ . 
\label{string}
\ee
Consider an arbitrary D p-brane
in the open bosonic string theory. The D p-brane longitudinal coordinates
can be defined at the open string boundaries{\footnote{$X^i_L(0,\tau )$ and
$X^i_R(\pi ,\tau )$ for $i = (0,1,2,\dots p-1)$
correspond respectively to the left and right ends of the open string.
Since the left and right modules are
commutant of each other, it is sufficient to consider a single module
of the string for the analysis.}},
$\s=0$ and $\s=\pi$, with Neumann condition.{\footnote{Here after, the
open string boundary is assumed to be implicit and we shall directly refer
to a D p-brane.}} 
In this picture, the world-volume coordinate fields $X^i$ are real 
and so its zero modes. Also the D p-brane
non-zero modes ($m\neq 0$) satisfy the
hermiticity condition: $\a^i_{-m}=(\a^i_m)^{\dag}$
which is evident from the mode expansion
(\ref{string}). The equal-time ($\tau$) canonical
commutation relation for the world-volume coordinates can be re-written
from that of the open string: 
\be
\Big [\ X^i(\s,\tau)\; ,\; P^j(\s',\tau)\ \Big ]\ = i\ g^{ij}\ \d (\s-\s') 
\ , \label{com}
\ee
where $P^i= (2\pi \ g_s\a')^{-1} \pr_{\tau}X^i$ 
represents the conjugate momenta and $(2\pi\ g_s\a')$ signifies the
D p-brane tension.
$g_{ij}$ denotes the induced metric on the world-volume. For simplicity,
we consider $g_{00}=-1$ and $g_{ij}=g\ \d_{ij}$ for $i\neq 0$.
Similarly, the commutation relation among the brane modes can be obtained 
at the string boundary $\pr\S$. The only non-vanishing commutator turns 
out to be the one among its non-zero momentum modes:
\be
\Big [\ \a^i_m\; ,\; \a^j_{m'}\ \Big ]\ = (2\a'\ m)\ 
g^{ij} \  \d_{m+m'}
\ . \label{ordcom}
\ee
The commutation relation implies
that the non-zero oscillator modes ($m\neq 0$)
can act as creation and anhilation operators, respectively,
for $m < 0$ and $m > 0$ in the Fock space. A careful analysis of the
non-zero modes towards its cyclic ordering (in $\tau$) among the
vertex operators can be performed. For instance, the identity
for two tachyon vertices with external momentum $k_1$ and $k_2$
turns out to be
\be
{\cv}_1(k_1,\tau_1)\ {\cv}_2(k_2,\tau_2)\ = {\cv}_2(k_2,\tau_2)
\ {\cv}_1(k_1,\tau_1)\ 
\cdot\exp{\left (\ 2i\pi\a'\ k_{1}\cdot k_{2}\ \ce (\tau_1-\tau_2)
\ \right )}
\ ,
\ee
where ${\ce}(\tau)$ is a constant and is not well defined at $\tau=0$.
Then for $n$-particles with external momentum $k_n$, 
the vertex operators $\cv_{n}(k_n)$ 
can be shown to be associated with a
phase factor:
\be
{\cv}_1(k_1,\tau_1)\dots {\cv}_n(k_n,\tau_n)\ 
\longrightarrow \ \exp \left ( {2i\pi\a'}
\sum_{n < n'}g^{ij}\ k_{ni} k_{n'j}\ {\ce}(\tau_n-\tau_{n'})\right )\;
{\cv}_1(k_1,\tau_1)\dots {\cv}_n(k_n,\tau_n) \ . \label{cs} 
\ee
However,
using the momentum conservation, the phase factor reduces to
an identity which is a requirement of the cyclic symmetry and leads to
an unitary description. 

\subsection{Noncommutative description on the world-volume}

Now consider a magnetic field ($B_{0i}=0$) on the world-volume of a D p-brane. 
The magnetic field can be seen to introduce non-perturbative correction to the
world-volume coordinates $X^i\ \rightarrow {\hat X}^i$ and makes
it noncommutative.

\sp

Naturally, the magnetic field significantly modifies the
the effective dynamics of a
D p-brane.{\footnote{The dynamics can be obtained from
the disk amplitude modulo closed string vertex with the
appropriate boundary conditions. In
an orthonormal moving frame \cite{kar98},
the action for a D p-brane simplifies
drastically. The locally flat background fields, in the frame-work, facilitates
the computation and leads to a general description for a
D p-brane dynamics with symmetric and antisymmetric 
curvatures.}}
The on-shell condition for the world-volume coordinates turns out to be:
\be
g_{ij}\ \pr_{\s}{\hat X}^j\ +\ {\bar B}_{ij}\ \pr_{\tau}{\hat X}^j 
\ {\Big |}_{\pr\S}
= 0 \ .
\label{condition}
\ee
In this case, the equal-time commutation relation for the conjugate
coordinates  satisfies (\ref{com}). However, the conjugate momenta receives
correction due to the magnetic field. The momenta can be given by
\be
{\hat P}^i\ =\ {1\over{2\pi\ g_s\a'}}\left [g^{ij}\ \pr_{\tau}{\hat X}_j\ 
+\ {\bar B}^{ij}\ \pr_{\s}{\hat X}_j \ \right ]\ . \label{momenta}
\ee
In presence of a magnetic field, the equal-time commutation relation 
for the world-volume coordinates turns out to be{\footnote{For instance,
the Dirac bracket for the open string coordinates at its
boundary has been obtained in ref.\cite{chu}.
We follow the notions in ref.\cite{kar99}.}}
\be
\Big [\ {\hat X}^i(\tau ) \; ,\; {\hat X}^j(\tau )\ \Big ] \
= \ \pm i\ (2\pi\a') \ \Theta_A^{ij} \ ,\label{ncom}
\ee
where $\Theta_A^{ij}$ is an antisymmetric matrix and is invertible. 
Explicitly{\footnote{In an orthonormal frame, $\Theta_A^{ij}$
can be locally mapped to a constant matrix.}}
\be
\Theta_A^{ij} = g_s\left ({1\over{g+{\bar B}}}{\bar B}{1\over{g -
{\bar B}}}\right )^{ij} \ .\label{parameter}
\ee
In fact, it is the antisymmetric property of an induced field 
that gives rise to a noncommutative description
on the D p-brane world-volume. The noncommutativity
of the world-volume coordinates in presence of a magnetic field is
a new phenomena and can be seen essentially due to its 
non-perturbative corrections. For instance, in absence of a magnetic field,
the r.h.s. in the commutation relation (\ref{ncom}) vanishes and implies
an ordinary geometry. It can be checked that any
perturbative correction (in $\Theta_A$) to the world-volume coordinates
can not change an ordinary geometry to a noncommutative one.

\section{Brane modes in presence of a magnetic field}

\subsection{Zero modes noncommutativity}

Since a D p-brane is defined at the boundary of an open string, 
the world-volume can be described by the string modes. 
The Fourier mode expansion for the coordinate fields $X^i(\tau )$ can
be obtained by drawing an analogy from the string mode expansion
(\ref{string}) which satisfies the modified
boundary condition (\ref{condition}). It becomes:
\be
{\hat X}^i(\tau )\ =\ {\hat x}^i\ + \ (2\a')\ {\hat p}^i\tau 
\ +\ i{\sqrt{2\a'}}\
\sum_{m\neq 0}{1\over{m}}\ {\hat\a}^i_{m}
\ e^{-im\tau} \ \cos\ m\s\ \Big |_{\pr\S}
\ . \label{brane}
\ee
Since the conjugate momenta (\ref{momenta}) is modified in presence of a 
magnetic field, a careful analysis can be performed 
\cite{kar98,kar99} to see that the induced
metric also receives correction. We recall \cite{seiberg} that the
corrections are obtained at the open string boundary.
As a result the effective metric in presence
of a magnetic field turns out to be:
\be
g^{ij}\ \longrightarrow \ \Theta_S^{ij} \ = 
\left ({1\over{g+{\bar B}}} g {1\over{g -
{\bar B}}}\right )^{ij} \ .\label{rmetric}
\ee
The canonical commutator (\ref{com}) in presence of a magnetic
field for the conjugate zero modes can be defined with the modified metric
(\ref{rmetric}). It becomes
\be
\Big [\ {\hat x}^i\; ,\; {\hat p}^j\ \Big ]
 = i\ \Theta_S^{ij}
\ . \label{com1}
\ee
Similarly, the commutation relation among the non-zero modes (\ref{ordcom})
in presence of a magnetic field can be rewritten as
\be
\Big [\ {\hat \a}^i_m\; ,\; {\hat \a}^j_{-m}\ \Big ]\ = (2\a'\ m)\
\Theta_S^{ij} \  
\ . \label{npcom}
\ee
However the equal-time canonical commutation relation (\ref{com}) in presence
of a magnetic field remains unaffected, since the corrections are incorporated
in the momenta ${\hat P}^i$. For a D p-brane, the non-zero modes are 
constrained at the open string boundary 
to yield ${\hat P}^i=g_s^{-1}{\hat p}^i$.
The commutator can be checked for the brane modes and
becomes:
\be
\Big [\ {\hat X}^i(\s,\tau)\; ,\; {\hat P}^j(\s',\tau)\ \Big ]
 = i\ g^{ij}\ \d (\s -\s')
\ , \label{com2}
\ee
where $\d (\s -\s')= \pi^{-1}\left ( 1 + \sum_{m\neq o}
\cos m\s\ \cos m\s'\ \right )$ defines at the string 
boundary.{\footnote{In general, $\d (\s -\s')$ can be seen to
be constrained by the oscillator modes, which allow $\s$ to take
one value in bulk apart from its boundary values.}}
The analysis in presence of a magnetic field  
(\ref{com2}) re-confirms the world-volume commutation 
relation obtained in eq.(\ref{ncom}).

\sp

Now the commutatation relation (\ref{ncom}) can be re-written, 
independently, in terms of its zero and non-zero modes (\ref{brane}).
Then the equal-time commutator reduces to the one among its zero modes
${\hat x}^i$. It can be given by
\bea
&&\Big [\ {\hat x}^i \; ,\; {\hat x}^j\ \Big ] =
\ \pm i\ (2\pi\a')\ \Theta_A^{ij} 
\nonumber \\
{\rm and}\qquad &&\Big [\ {\hat p}^i \; ,\; {\hat p}^j\ \Big ] =\ 0 \ .
\label{magcommutator}
\eea
The non-vanishing of the commutation relation 
for the constant coordinate modes can be understood
in terms of its non-perturbative correction.
Since a magnetic field
defines a noncommutative world-volume (\ref{ncom}),
it is the constant coordinate modes, $x^i$, which inherit the effective
noncommutativity (\ref{magcommutator}).
The constant (canonical) momentum modes commute in this channel
as they do not receive any correction.

\sp

On the other hand, the non-zero modes 
on the world-volume are modified in presence of a magnetic field
(\ref{npcom}). They are redundant to a noncommutative
description though their role is vital towards an UV-renormalization
in an ordinary theory. 

\subsection{Non-local phase factor}

In this section, we compute the phase factor
associated with the constant coordinate modes in a noncommutative description.
A Weyl ordering prescription can be utilized to
map an ordinary scalar field $\phi(x)$ to 
$\Phi({\hat x})$ in an ordered space. Since the noncommutative world-volume
is defined with $({\hat x}^i-x^i)\neq 0$, a scalar field can be given by
\be
\Phi({\hat x}) =\ {1\over{(2\pi)^{p}}}
\ \int \ d^{p}x\; d^{p} {\hat p}\
\; e^{i{\hat p}_i \cdot ({\hat x}^i-x^i)}\ \phi(x) \ ,
\label{hom}
\ee
where a note on the order of the homomorphic map 
$\phi(x)\rightarrow \phi({\hat p})\rightarrow 
\Phi({\hat x})$ is implicit in eq.(\ref{hom}).
Then, a point-wise multiplication is replaced by the point-less 
(noncommutative) $\star$-product.
The scalar interaction term for $n$-number of
particles can be given in its momentum space:
\bea
\Phi({\hat x})\star \ \dots \ \star \Phi({\hat x})
&=& {1\over{(2 \pi)^{p/2}}} \int \ d^{p}{\hat p}_1\dots d^{\hat p}{p}_n
\nonumber\\
&&\qquad\qquad\qquad\quad
\cdot e^{i{p}_1\cdot {\hat x}}\ \phi({\hat p}_1)\
\dots \ e^{i{\hat p}_n\cdot {\hat x}}\ \phi({\hat p}_n)
\ . \label{star1}
\eea
A simplification of the integrand can be performed with the help of the
commutation relation (\ref{magcommutator}). Then the $n$-particle scalar
interaction (\ref{star1}) reduces to:
\bea
\left [\Phi({\hat x})\right ]^n &=&
{1\over{(2 \pi)^{p/2}}} \int \ d^{p}{\hat p}_1\dots d^{p}{\hat p}_n\;
\cdot\exp \left({\ i \sum_n {\hat p}_n\cdot {\hat x}
\pm \ i\sum_{n<n'}\ {\hat p}_{ni} \Theta_A^{ij} {\hat p}_{n'j}}\right )
\nonumber\\
&& \qquad\qquad\qquad\qquad\qquad\qquad\qquad\qquad
\Big [\ \phi({\hat p}_1)\ \dots \ \phi({\hat p}_n)\ \Big ]
\ .
\eea
The integrand confirms that the $n$-particle interaction
vertex is associated with a phase factor in a noncommutative description.
It implies that the internal momentum ${\hat p}$ plays 
significant role in a noncommutative theory.
The momentum conservation condition
further simplifies the $n$-particle interactions
and the integrand correspond precisely
to a vertex $\cv_{\phi}({\hat p}_1,\dots \ ,
{\hat p}_n)$ in the momentum space. It is
given by
\be
\cv_{n\phi}({\hat p}_1,\dots \ ,{\hat p}_n)\
=\ \d^{(p)}\left ( \sum_n {\hat p}_n\right )
\ \exp \left (\ \pm i\sum_{n<n'}\ {\hat p}_{ni}\Theta_A^{ij}
{\hat p}_{n'j}\right )
\cdot \Big [\phi({\hat p}_1)\dots \phi({\hat p}_n)\Big ]
\ . \label{vertex2}
\ee
It shows that the (scalar) interaction vertex on a noncommutative world-volume 
is associated with a non-trivial phase factor in addition to the one
obtained (\ref{cs}) due to the cyclic property. 
Thus a non-local quantum field theory may be expressed in
terms of a phase factor due to the constant modes and an ordinary
quantum field theory, which can be seen to be UV-renormalized.
Then the scattering amplitude in a noncommutative
theory can be given by a combination of its oscillatory phase defined
with the internal momentum $p$ and an UV renormalized QFT for the interacting 
particles with external momentum $k$. Schematically:
\be
{\ca}\left ( k_n,{\hat p}_n; k_{n'},{\hat p}_{n'}\right )_{NQFT}\ 
\equiv\ \exp \ \left (\pm i\sum_{n<n'}
\ {\hat p}_{ni}\Theta_A^{ij}{\hat p}_{n'j} \right )\ \cdot
{\bar {\ca}}\left ({k_n,k_{n'}}\right )_{QFT} \ , \label{relation}
\ee
where an appropriate momentum conservation holds at the interacting vertices.
The expression (\ref{relation}) is supported by the fact that the zero modes
describe the ground state of a closed string and are massless.
The wave function for a
D p-brane in its ground state can be seen to be degenerate though
energy level is not. Indeed, the multi-valued ground state wave function
gives rise to a non-local description \cite{kar99}. Then the task reduces 
to the computation of the non-local phase factor (\ref{relation}). In fact, the
phase dominates over the usual behavior of momentum in
the theory and softens the loop divergences. 

\sp

Finally, we conclude the section with a note that
the zero modes on the world-volume receive non-perturbative correction
in presence of a magnetic field. They describe a
noncommutative theory and contains its ordinary QFT counterpart.
The fact lead to a conjecture that a noncommutative theory is
renormalizable if its ordinary counterpart does.

\subsection{Field theoretic limit ($\a'\ \rightarrow 0$)}

In the previous section, we noticed that a D p-brane center of mass
coordinates decide the nature of its world-volume geometry.
In presence of a magnetic field, the constant modes ${\hat x}^i$ receive
geometric corrections (in $\Theta_A$) and the world-volume turns out to be
noncommutative. On the other hand, the non-zero modes are associated
with the stringy corrections (in $\a'$) and do not play any
role towards its world-volume geometric corrections. Thus, these two
corrections are independent of each other.

\sp

In a field theoretic limit, $\a'\rightarrow 0$, the non-zero modes
are dropped out and the world-volume retains its noncommutative
description. A priori, the UV-renormalization seems to be not guaranteed
in the field theoretic frame-work. However a close note on the world-volume
noncommutative description (\ref{ncom}), yields an uncertainty relation:
\be
\Delta {\hat x}^i\cdot\Delta {\hat x}^j\ \geq\ 
{1\over2}\left |{\Theta_A^{ij}}\right | \ .
\label{uncertainty2}
\ee
It confirms a finite value to the, naive, short-distance divergence
by assigning  a critical cut-off value $(\Delta {\hat x}^i)_{\rm crit}\neq 0$
for $i\neq j$. As a result, a minimum length scale is introduced into a
noncommutative theory which takes care the UV-divergences. For
instance, keeping a particle position (${\hat x}^i$) fixed, the uncertainty
relation (\ref{uncertainty2}) would lead to non-local description in
the ($i\neq j$)-directions.
Since in absence of either ($\a'$ or $\Theta_A$) corrections, the
space contains UV-divergences, these two parameters seem to play an identical
role towards the short-distance features of a theory. Geometrically,
$\a'$ and $\Theta_A$ show their differences, nevertheless
their UV-features can be seen to be identical in section 4.
For instance, the $\a'$-correction introduces
infinite number of particles in the spectrum and defines a string of
length ${\sqrt{2\a'}}$ where as the $\Theta_A$-correction gives rise to
a large density of particles in the ground state
(zero modes) and makes it non-local. As explained, the latter 
phenomena introduces the notion of a noncommutative geometry on a D p-brane
world-volume.

\section{UV and IR phenomena}

\subsection{Cyclic symmetry in presence of a magnetic field}

We begin this section by considering a two point correlator \cite{kar98}
for the open string vertex operators evaluated at the string boundary. 
Effectively, it describes a propagator on a D p-brane world-volume.
The Neumann propagator in a matrix notation can be written as
\be
{\langle}\ {\hat X}^i(\tau){\hat X}^j(\tau ')\ {\rangle}\
=\ 4\pi\a' \ \Theta_S^{ij} \ \ln (\tau -\tau' )
\ - \ {i\pi\a'}\ \Theta_A^{ij}\ {\ce } (\tau -\tau' ) \ ,
\label{prop}
\ee
where ${\ce } (\tau)$ introduces
the notion of $\tau$-ordered space on the string world-sheet.
In the alternate picture, the noncommutative description
on the world-volume{\footnote{It is associated with an antisymmetric exchange
of $i\leftrightarrow j$  indices}} (\ref{ncom}) can be re-written in terms
$\tau$-ordering at the string boundary.
An appropriate combination of the propagators in the ordered space
for the world-volume operators ${\hat X}^i(\tau )$ turns out to 
be identical. It can be given by
\be
T\ \Big [\ {\hat X}^i(\tau ) {\hat X}^j(\tau' )\ - \ {\hat X}^i(\tau' ) 
{\hat X}^j(\tau )\ \Big ]\
= \ i\ (2\pi\a')\ \left ( 2\ \Theta_S^{ij}\  -\ \Theta_A^{ij} \right )\ 
\ce (\tau-\tau' ) \ . \label{time}
\ee
Now, interchanging the notion of time-ordering with the antisymmetric
exchange of brane coordinates, one can define the limit 
$\tau\rightarrow \tau'$. Then, with a little subtleties, one can obtain
the equal-time world-volume commutation relations (\ref{ncom}) 
from the time-ordered combination (\ref{time}). 
The analysis implies that an ordered space in presence of a B-field
can be equivalently described by a noncommutative space.
The antisymmetrization among the world-volume operators 
${\hat X}^i(\tau)$ naturally forbids the non-zero modes commutation
relation (\ref{npcom}). The analysis further assures the relevance of
zero coordinate modes ${\hat x}^i$ towards a noncommutative geometry. 
A priori the result may appear surprising, since it is the strong
magnetic field (large B) limit where the zero modes play a leading
role towards the world-volume geometry. However it works, since
the usual role of non-zero modes{\footnote{They play
a crucial role towards the UV-renormalization in an ordinary
theory.}} is auxiliary in a noncommutative description.{\footnote{Alternately,
the noncommutative description defined by the equal-time commutation relation
(\ref{ncom}) can be checked for the brane modes (\ref{brane}). Formally
the modified metric $\Theta_S^{ij}$ on the r.h.s. of the commutator
can be seen to be associated with a divergent piece
($- 2\a' \ln m$) due to the infinite tower of non-zero modes. 
In an ordinary theory
the $\zeta$-function regularization takes care of the divergence.
On the other hand, a noncommutative description does not permit any symmetric
(in $i\leftrightarrow j$) term on the r.h.s. of its world-volume
commutation relation (\ref{ncom})
and hence forbids the otherwise divergent piece.}} 
It implies that the analysis for a large
B-field may be generalized to any $B\neq 0$. In particular, the large B limit
is an important domain where the theory turns out to be topological and
holographic idea can be studied there.

\subsection{Origin of virtual particles and  IR-singularities} 

Until now, we have been discussing the D p-brane modes 
in an open string channel (t-channel). 
We learned that in a noncommutative description, the uncertainty relation
(\ref{uncertainty2}), among the constant coordinate modes ${\hat x}^i$,
takes care the  otherwise short-distance divergences. 
In addition, these modes
are non-propagating ones which is evident from their phase contribution
(\ref{vertex2}). In the open string channel,
the constant modes (\ref{brane}) can be easily 
identified with that of a closed string and are massless. Since the
constant momentum modes $p^i$ commute (\ref{magcommutator}), the closed
string zero modes do not possess a lower cut-off limit in its
internal momentum. 
As a result, the loop integral for the zero modes 
exhibit the usual IR singularities on the  D p-brane world-volume.

\sp

To be explicit, consider $n$ (massive) particle interaction
vertex{\footnote{In case of scalar
particles, the interaction vertex has been obtained in eq.(\ref{vertex2}).}}
(for $n'<n$) with external
momenta $(k_1,\dots ,k_{n'},\dots , k_{n})$ in a noncommutative theory.
In a point-splitting prescription, the interaction vertex $\cv_n$ splits
into two vertices, each consisting of $n'$ and $(n-n')$ interacting
particles, and are connected by the matrix correlator
\be
{\left\langle \; {\hat\phi}_0^i(-p)\ {\hat\phi}_0^j(p)\; \right\rangle }\ =
\ \pm (\pi\a')\; i \Theta_A^{ij}\ .
\ee
Here ${\hat\phi}_0^i$ denote the closed string 
zero modes and represent 
virtual particles in the momentum space.
The $n$-particle vertex can be illustrated by a point-splitting 
prescription:
\be
\cv(k_1,\dots , k_{n'},k_{n'+1},\dots , k_{n})
\ \rightarrow\ \cv_{1i}(k_1,\dots ,k_{n'})
\; (i\  \Theta_A^{ij})\ \cv_{2j}(k_{n'+1},\dots ,k_{n}) \ ,
\label{split}
\ee
where the momentum conservation and the cyclic symmetry are
preserved at both the vertices. 
Intuitively, the point-splitting prescription 
is identical to the stringy phenomenon involving different channels. 
Alternately, the point-splitting prescription (\ref{split}) can be 
described by two independent vertices with momentum conservation 
condition at each of them:
$${\tilde \cv}_1(k_1,\dots ,k_{n'},-{p})
\cdot  {\tilde \cv}_2(p,k_{n'+1},\dots ,k_{n}) \ .$$
For scalar particles, the vertices can be expressed explicitly:
\bea
&& {\tilde \cv}_{1\phi} 
= \d^{(p)}\left (k_1+\dots +k_{n'}-{p}\right ) \ \cdot\exp \left (
\pm i\ p_i\Theta_A^{ij}p_j\right )
\ \Big [\phi(k_1)\dots \phi(k_{n'}){\hat\phi}_0(-{p})\ \Big ]
\nonumber \\
{\rm and}&&{\tilde \cv}_{2\phi} 
= \d^{(p)}\left (k_{n'+1}+\dots +k_{n}+p\right ) \cdot\exp \left (
\pm i \ p_i\Theta_A^{ij}p_j \right ) \
\Big [\phi(k_{n'+1})\dots \phi(k_{n}) {\hat\phi}_0(p)\Big ] \ .
\label{2vertex}
\eea
The momentum conservation at each vertex yields:
\bea
&& {\tilde \cv}_{1\phi}\
= \ \exp \left (
\pm i\ k \cdot p_{nc}\right )
\ \Big [\ \phi(k_1)\dots \phi(k_{n'}){\hat\phi}_0(-{p})\ \Big ]
\nonumber \\
{\rm and}\:
&& {\tilde \cv}_{2\phi}\ = \ \exp \left (
\mp i \ k'\cdot p_{nc} \right ) \ 
\Big [\ \phi(k_{n'+1})\dots \phi(k_{n}) {\hat\phi}_0(p)\ \Big ] \ ,
\label{3vertex}
\eea
where $p_{nc}^i=\Theta_A^{ij}p_j$ defines a noncommutative momenta.
There $k$ and $k'$ denote the total momentum due to the external
particles, respectively, 
with ${\tilde \cv}_{1\phi}$ and ${\tilde \cv}_{2\phi}$. 
Thus in a noncommutative description, the $n$-particle vertex
can be described by two independent vertices with 
$(n'+1)$ and $(n-n'+1)$ number of interacting particles. 
The additional particles at both the vertices are essentially the
closed string zero modes and give rise to the IR poles in the amplitude
even for the massive external particles. The phenomenon has been addressed 
\cite{minwalla}-\cite{rajaraman} 
explicitly for the massive scalar $\Phi^4({\hat x})-$interaction 
in four dimensions. In fact, the t-channel analysis leads
to a non-planar Feynman graphs where the oscillatory phase 
(\ref{relation}) dominates and softens the otherwise divergences. The
limiting value to the effective momentum, in t-channel, turns out to 
be: $\L_{t}\ \equiv\ p_{nc}.$
In the limit $\Theta_A\rightarrow 0$, the effective cut-off $\L_{t}
\rightarrow \infty$ and corresponds to the one in an ordinary theory. 
However, the cut-off $\L_{t}$ remains finite for the non-planar
loop graphs.

\subsection{Effective IR cut-off}

Under a T-duality symmetry, the one-loop amplitude in a t-channel can be
transformed to a tree amplitude in closed string theory (s-channel).
In absence of a magnetic field, the Neumann matrix propagator (\ref{prop})
becomes diagonal and contains a short-distance loop divergence in the limit
$\tau\rightarrow\tau'$. It corresponds to an UV-divergence
in the t-channel and can be regulated by defining a limiting value
in a standard prescription. The UV cut-off value is known to be interpreted
as the IR limit in its dual (s-) channel. Thus in a string frame-work,
very different phenomena can be seen to occur in different channels.
A close look in a noncommutative theory reveals that some phenomena
resemble to that of strings.

\sp

In presence of a magnetic field, the Neumann matrix
(\ref{prop}) contains off-diagonal propagators in addition to the usual
diagonal ones.
In the limit $\tau\rightarrow\tau'$, the underlying world-sheet theory
contains a loop diagram. However, the zero modes describing the
noncommutative geometry on the world-volume are free from loop
divergences and the matrix propagator can be seen to be a constant:
$ {\langle}\ {\hat x}^i{\hat x}^j\ {\rangle}\
=\ \pm \ {i\pi\a'}\ \Theta_A^{ij}.$
Interestingly, the propagator for the zero modes can
be illustrated by the non-planar Feynman graph \cite{filk}.

\sp

Since a duality transformation can be seen to
exchange the winding modes with the momentum ones, 
the world-sheet coordinates ($\s \leftrightarrow \tau$) interchange
their roles. In addition to it, the Dirichlet boundary conditions are
exchanged with the Neumann. Under a duality, the D ${\tilde p}$-brane
world-volume does not possess any magnetic field. On the other hand, the
magnetic field lies in the transverse space to the brane.
Thus in a closed string channel, the commutation relation among the
constant modes turns out to be
\bea
&&\Big [\ {\tilde x}^i \; ,\; {\tilde x}^j\ \Big ] =\ 0 
\nonumber \\
{\rm and}\qquad &&\Big [\ {\tilde p}^i \; ,\; {\tilde p}^j\ \Big ] =\ 
\pm i\ (2\pi\a')\ \Theta_A^{ij} \ .
\label{magcommutator2}
\eea
The commutator for the zero momentum modes
implies an uncertainty relation among them.
Thus the zero 
momentum mode ${\tilde p}$ is bounded from below by the matrix 
parameter $\Theta_A^{ij}$. In this case, though the closed string zero
modes are massless, the loop integral turns out to be free from any 
IR singularities. The IR cut-off value for the internal momentum 
${\tilde p}$, in s-channel, turns out to be:
$\L_{s} =1/{\sqrt{(\pi\a') \Theta_A}},$
where $\Theta_A$ denotes a typical eigen-value for the matrix 
$(\Theta_A)_{ij}$. The limit $\Theta_A\rightarrow 0$ corresponds to
planar Feynman graphs and obviously associated with the poles. 
Since the resulting amplitude in a scattering phenomena is obtained by
summing over its channel amplitudes, it implies that a
noncommutative theory always contain an
ordinary counterpart which is in agreement with the 
expression (\ref{relation}).

\section{Discussions}

We have presented some similarities between a noncommutative
theory and an open string theory in a non-perturbative frame-work of 
a D p-brane. We learned that the zero modes in the theory are completely
responsible for a noncommutative description on a D p-brane world-volume. 
Since in a field theoretic limit ($\a'\rightarrow 0$), the non-zero modes
decouple, the world-volume of a D p-brane precisely describes a NQFT. 
This, in fact, gives rise to a non-local QFT. Thus the non-locality is 
essentially due to the zero modes in the theory. Since the zero modes
are non-propagating fields, their contribution is expressed 
as a non-trivial phase factor in a momentum space. The phase obtained
is defined with the internal momenta of the zero modes. In a
scattering phenomena ($i.e.$ with external momenta of interacting particles)
the momentum conservation allows an interplay between the internal and
external momenta and redefines the phase. The IR-domain 
was analyzed to conclude that the virtual particles correspond to
the closed string zero modes and play a significant role in a noncommutative
theory.

\sp

On the other hand, the non-zero modes are associated with the $\a'$
perturbative corrections. It was argued that the $\Theta_A$ and $\a'$
play an identical role towards the renormalizations.
In other words, the role of zero modes in
a noncommutative theory is identical to that of the non-zero modes
in a string theory. Both the modes satisfy identical commutation relations.
A point-splitting prescription in the former, apparently seems to yield
scattering processes identical to that in the T-duals in string theory.
In addition, a noncommutative description correspond to a large number 
of particles (which is transparent in the large B limit)
in the spectrum similar to
that of string. A priori, there seems to be an ambiguity, at this
level, since the string spectrum consists of massive particles where
as the zero modes are mass-less. However, the ambiguity may be resolved
by generalizing the large B limit to any $B\neq 0$, where presumably the
gravity would supplement for mass to the zero modes. It is in
agreement with the fact of an IR cut-off for the zero modes in the theory. 
In the context it remains to understand, if holography has a role to 
play in a noncommutative theory.

\sp

\sp

\noindent {\Large\bf Acknowledgments}

\vspace{.1in}

I am grateful to the members of the string group, in the Institute, for their
helpful comments during an informal presentation of this work at its early 
stage. The work is supported by the Swedish Natural Science Research Council.

\vfil\eject 

\def\anp{Ann. of Phys.}
\def\prl{Phys. Rev. Lett.}
\def\prd#1{{Phys. Rev.} {\bf D#1}}
\def\plb#1{{Phys. Lett.} {\bf B#1}}
\def\npb#1{{Nucl. Phys.} {\bf B#1}}
\def\mpl#1{{Mod. Phys. Lett} {\bf A#1}}
\def\ijmpa#1{{Int. J. Mod. Phys.} {\bf A#1}}
\def\rmp#1{{Rev. Mod. Phys.} {\bf 68#1}}


\end{document}